\documentclass[a4paper,aps,prl,twocolumn,preprintnumbers,showpacs,superscriptaddress,nofootinbib]{revtex4}
\usepackage{graphics,graphicx,epsfig}
\usepackage{amsfonts,amsmath,amssymb}
\usepackage{psfrag,verbatim}
\newcommand{\be}{\begin{equation}}
\newcommand{\ee}{\end{equation}}
\newcommand{\beq}{\begin{equation}}
\newcommand{\eeq}{\end{equation}}
\newcommand{\ba}{\begin{eqnarray}}
\newcommand{\ea}{\end{eqnarray}}
\newcommand{\bea}{\begin{eqnarray}}
\newcommand{\eea}{\end{eqnarray}}

\def\cE {{\mathcal{E}}}
\def\cF {{\mathcal{F}}}

\def\cN {{\mathcal{N}}}
\def\cO {{\mathcal{O}}}
\def\cP {{\mathcal{P}}}

\def\cS {{\mathcal{S}}}

\def\a {\alpha}

\def\e {\epsilon}

\def\s {\sigma}

\def\D {\Delta}

\begin{document}

%%%%%%%%%%%%%%%%%%%%%%%%%%%%%%%%%%%%%%%%%%%%%%%%%%%%%%%%%%%%%%%%%%%%%%%%%%%%%%%%%%%%%%%%%%%%%%%%%%%%%%%%%%%%%
%%%%%%%%%%%%%%%%%%%%%%%%%%%%%%%%%%%%%%%%%%%%%%%%%%%%%%%%%%%%%%%%%%%%%%%%%%%%%%%%%%%%%%%%%%%%%%%%%%%%%%%%%%%%%
%%%%%%%%%%%%%%%%%%%%%%%%%%%%%%%%%%%%%%%%%%%%%%%%%%%%%%%%%%%%%%%%%%%%%%%%%%%%%%%%%%%%%%%%%%%%%%%%%%%%%%%%%%%%%
\title{Cavitation effects on the confinement/deconfinement transition}
\author{Alex Buchel}
%\email{abuchel@perimeterinstitute.ca}
\affiliation{Perimeter Institute for Theoretical Physics, Waterloo, Ontario N2J 2W9, Canada}
\affiliation{Department of Applied Mathematics, University of Western Ontario
London, Ontario N6A 5B7, Canada}
\author{Xi\'an O. Camanho}
%\email{xian.otero@rai.usc.es}
\affiliation{Department of Particle Physics and IGFAE, University of Santiago de Compostela\\ E-15782 Santiago de Compostela, Spain}
\author{Jos\'e D. Edelstein}
%\email{jose.edelstein@usc.es}
\affiliation{Department of Particle Physics and IGFAE, University of Santiago de Compostela\\ E-15782 Santiago de Compostela, Spain}
\affiliation{Centro de Estudios Cient\'{\i}ficos CECs, Av. Arturo Prat 514, Valdivia, Chile}
\pacs{11.25.Tq, 12.38.Mh, 25.75.Nq}
\preprint{UWO-TH-13/4}

\begin{abstract}
Cavitation is a process where the viscous terms in a relativistic fluid result in reducing the effective pressure, thus facilitating the nucleation of bubbles of a stable phase. The effect is particularly pronounced in the vicinity of a (weak) first-order phase transition. We use the holographic correspondence to study cavitation in a strongly coupled planar cascading gauge theory plasma close to the confinement/deconfinement phase transition. While in this particular model the shift of the deconfinement temperature due to cavitation does not exceed $5\%$, we speculate that cavitation might be important near the QCD critical point.
\end{abstract}
\maketitle
%%%%%%%%%%%%%%%%%%%%%%%%%%%%%%%%%%%%%%%%%%%%%%%%%%%%%%%%%%%%%%%%%%%%%%%%%%%%%%%%%%%%%%%%%%%%%%%%%%%%%%%%%%%%%
%%%%%%%%%%%%%%%%%%%%%%%%%%%%%%%%%%%%%%%%%%%%%%%%%%%%%%%%%%%%%%%%%%%%%%%%%%%%%%%%%%%%%%%%%%%%%%%%%%%%%%%%%%%%%
%%%%%%%%%%%%%%%%%%%%%%%%%%%%%%%%%%%%%%%%%%%%%%%%%%%%%%%%%%%%%%%%%%%%%%%%%%%%%%%%%%%%%%%%%%%%%%%%%%%%%%%%%%%%%

%%%%%%%%%%%%%%%%%%%%%%%%%%%%%%%%%%%%%%%%%%%%%%%%%%%%%%%%%%%%%%%%%%%%%%%%%%%%%%%%%%%%%%%%%%%%%%%%%%%%%%%%%%%%%
\noindent
\textit{Introduction.---}
%%%%%%%%%%%%%%%%%%%%%%%%%%%%%%%%%%%%%%%%%%%%%%%%%%%%%%%%%%%%%%%%%%%%%%%%%%%%%%%%%%%%%%%%%%%%%%%%%%%%%%%%%%%%%
Hydrodynamics is a universal framework to describe strongly coupled systems at energy scales much lower than their characteristic microscopic scales (masses, temperature, etc). A noteworthy example, on which we shall lay our focus, is the plasma of quarks and gluons produced in relativistic heavy ion collisions. The basic hydrodynamic equation is that of the conservation of the stress-energy tensor
\be
\nabla_\mu T^{\mu\nu}=0 \,.
\label{const}
\ee
For an ideal  relativistic fluid the stress-energy tensor takes the form 
\be
T^{\mu\nu}_{ideal}=\cE\ u^\mu u^\nu +\cP\ \Delta^{\mu\nu}\,,
\label{tideal}
\ee
where $\cE$ and $\cP$ are the energy density and pressure,
$$
\Delta^{\mu\nu}=g^{\mu\nu}+u^\mu u^\nu\,, 
$$
and $u^\mu$ is the fluid four-velocity, normalized so that $u_\mu u^\mu=-1$. The leading viscous corrections are parameterized by the fluid shear $\eta$ and bulk $\zeta$ transport coefficients in the viscous tensor $\Pi^{\mu\nu}$:
\bea
&&T^{\mu\nu}=T^{\mu\nu}_{ideal}+\Pi^{\mu\nu}\,,
\label{ns} \\ [0.7em]
&&\Pi^{\mu\nu}=-\eta\ \s^{\mu\nu}-\zeta\;\nabla u\ \Delta^{\mu\nu}\,,
\eea
where $\nabla u \equiv \nabla_\a u^\a$, and we have adopted 
\be
\sigma^{\mu\nu} = \left(\D^{\mu\lambda} \nabla_\lambda u^\nu + \D^{\nu\lambda} \nabla_\lambda u^\mu\right)
- \frac{2}{3}\;\nabla u\ \D^{\mu\nu}\,.
\ee
Assuming that the relevant microscopic scale is the temperature, the leading hydrodynamic approximation is valid provided 
\be
\frac{|\nabla_{\mu} u^{\nu}|}{T}\ll 1\,,
\label{upper}
\ee 
otherwise, higher-order gradients (typically infinitely many of them) must be included \cite{Muronga:2001zk,Muronga:2003ta,Baier:2007ix}.

It is easy to see that viscous terms tend to reduce the pressure\footnote{For an expanding quark-gluon plasma fireball, it was argued in \cite{Rajagopal2010} that this may trigger cavitation, releasing the bunch of droplets that are required by models of statistical hadronization.} \cite{Rajagopal2010}. For example, for fluids comoving in an expanding background such as an FRW metric,
\be
ds_4^2=-dt^2+a(t)^2 (d\vec{x})^2\,,
\ee 
we find 
\be
\s^{\mu\nu}\big|_{\rm FRW}=0\,,\qquad \nabla u\big|_{\rm FRW} =3\frac {\dot{a}}{a}\,,
\ee
resulting in an isotropic reduced effective pressure
\be
\cP^{\rm eff} = \cP - \zeta\;\nabla u\,.
\label{pfrw}
\ee
In the case of a boost invariant fluid expansion, the pressure is no longer isotropic \cite{Muronga:2003ta}:
\be
\cP_{\perp}^{\rm eff}=\cP+\frac{2\eta-3\zeta}{3\tau}\,,\qquad \cP_\xi^{\rm eff}=\cP-\frac{4\eta+3\zeta}{3\tau}\,,
\ee
where $ _\perp$ and $ _\xi$ are the transverse and longitudinal directions of the boost invariant expansion\footnote{Such expansion is conveniently described changing variables from $(t,z)$ to $(\tau,\xi)$: $\tau=\sqrt{t^2-z^2}$, $\xi={\rm arctanh}\frac zt$.}, and $\tau$ is the proper time.  Notice that in this case the {\it spatially averaged} pressure, $\left(\frac 23 P_{\perp}+\frac 13 P_\xi\right)$ still takes the form (\ref{pfrw}). In what follows, we take (\ref{pfrw}) as a generic expression for the effective pressure.

~

%%%%%%%%%%%%%%%%%%%%%%%%%%%%%%%%%%%%%%%%%%%%%%%%%%%%%%%%%%%%%%%%%%%%%%%%%%%%%%%%%%%%%%%%%%%%%%%%%%%%%%%%%%%%%
\noindent
\textit{Cavitation and first-order phase transitions.---}
%%%%%%%%%%%%%%%%%%%%%%%%%%%%%%%%%%%%%%%%%%%%%%%%%%%%%%%%%%%%%%%%%%%%%%%%%%%%%%%%%%%%%%%%%%%%%%%%%%%%%%%%%%%%%
Consider now a system which, in thermal equilibrium, can exist in one of the two phases $A$ or $B$. A first-order phase transition between these phases implies the existence of a critical temperature $T_c$, such that $\cP_A> \cP_B$ for $T>T_c$, and $\cP_A< \cP_B$ otherwise. The phase with the higher pressure is thermodynamically favored, and the transition at $T=T_c$ proceeds through nucleation of bubbles of the stable phase. If the system flows, the relevant pressure determining the stability of a phase is the effective one:
\be
\cP_{A/B}^{\rm eff}=\cP_{A/B}-\zeta_{A/B}\;\nabla u\,.
\ee
We use the first law of thermodynamics, $d\cF=-d\cP=-\cS\ dT$, to write, close to $T_c$,
\be
\cP_{A/B}=\cP_c+\cS_{A/B}\, (T-T_c)+\cO\left((T-T_c)^2\right)\,,
\label{pAB}
\ee
where $\cS_{A/B}$ are the entropy densities of the corresponding phases. Thus, viscous hydrodynamics effects would shift the transition temperature according to 
\be
\frac{|\delta T_c|}{T_c}\ \sim\ \frac{|\zeta_A-\zeta_B|}{|\cS_A-\cS_B|}\ \frac{|\nabla u|}{T_c} \ \lesssim\ \frac{|\zeta_A-\zeta_B|}{|\cS_A-\cS_B|}\,,
\label{big}
\ee
where the upper bound is enforced from the consistency of truncating hydrodynamics at the first order in the velocity gradients, see (\ref{upper}). Notice that cavitation affects the transition temperature the more weakly the first-order transition (the smaller the difference between $\cS_{A/B}$) is, and the larger the bulk viscosity difference of the two phases at $T_c$. 

Ideally, we would like to evaluate (\ref{big}) for  QCD close to confinement/deconfinement transition.\footnote{In the context of the production of quark-gluon plasma by relativistic heavy ion collisions, we might remove the absolute values in (\ref{big}) and, $\nabla u$ being positive, it will lead to an increase in the temperature. This means that hadronization, if driven by cavitation, might start earlier than naively expected.} While the recent lattice results  provide a reliable equation of state \cite{Borsanyi:2010cj} (at least at vanishing baryon chemical potential), rather than doing it from first principles, one has to rely on various models to evaluate transport coefficients of gauge theory plasma at strong coupling \cite{Policastro:2001yc,Buchel:2008vz,Karsch:2007jc,Buchel:2007mf,Buchel:2009mf}. In what follows we present the first self-consistent estimate of (\ref{big}) for a strongly coupled gauge theory plasma. 

~

%%%%%%%%%%%%%%%%%%%%%%%%%%%%%%%%%%%%%%%%%%%%%%%%%%%%%%%%%%%%%%%%%%%%%%%%%%%%%%%%%%%%%%%%%%%%%%%%%%%%%%%%%%%%%
\noindent
\textit{Cascading gauge theory.---}
%%%%%%%%%%%%%%%%%%%%%%%%%%%%%%%%%%%%%%%%%%%%%%%%%%%%%%%%%%%%%%%%%%%%%%%%%%%%%%%%%%%%%%%%%%%%%%%%%%%%%%%%%%%%%
Consider $\cN=1$ four-dimensional supersymmetric $SU(K+P)\times SU(K)$ gauge theory with two chiral superfields $A_1, A_2$ in the $(K+P,\overline{K})$ representation, and two fields $B_1, B_2$ in the $(\overline{K+P},K)$ \cite{ks}. This gauge theory has two gauge couplings $g_1, g_2$ associated with the two gauge group factors,  and a quartic superpotential
\be
W\sim {\rm tr} \left(A_i B_j A_kB_\ell\right)\e^{ik}\e^{j\ell}\,.
\ee
The theory is not conformal, and develops a strong coupling scale $\Lambda$ through dimensional transmutation of the gauge couplings. In the UV/IR it undergoes the {\it cascade} of Seiberg dualities \cite{sd} with $K\to K\pm P$. The net result of the duality cascade is that the rank $K$ of the theory becomes dependent on the scale $E$ at which the theory is probed \cite{b}:
\be
K\to K_{\rm eff}(E)\approx 2P^2 \ln \frac{E}{\Lambda}\,,\qquad E\gg \Lambda\,.
\ee
While not  QCD, the theory shares some of  the IR features of the latter: when $K$ is an integer multiple of $P$, the cascade ends in the IR with $SU(P)$ supersymmetric Yang-Mills theory  which confines with spontaneous breaking of the chiral symmetry.

Cascading gauge theory is always strongly coupled in the UV. In the planar limit and for large 't Hooft coupling of the IR $SU(p)$ factor, the theory is strongly coupled along its full RG flow, and thus can be consistently studied using its holographic dual \cite{ks}. We focus on the cascading gauge theory in the regime where the holographic description is reliable.

Thermodynamics of the cascading gauge theory plasma has been studied extensively in the past \cite{kt3,hyd3,Buchel:2010wp}: it simultaneously undergoes (first-order) confinement and chiral symmetry breaking at $T_c=0.6141111(3) \Lambda$. Furthermore, the deconfined phase becomes unstable towards spontaneous development of a chiral condensate at a slightly lower temperature, $T_{\chi sb}=0.882503(0) T_c$. Finally, at $T_u=0.8749(0) T_c$, the deconfined phase of the theory approaches a critical point with a divergent specific heat \cite{Buchel:2009mf}.

The shear viscosity of the plasma is universal for all phases and at all temperatures \cite{bl},
\be
\frac{\eta}{\cS}=\frac{1}{4\pi}\,.
\ee 
The bulk viscosity of the theory is technically difficult to compute --- so far it is known only to the fourth order in the high temperature expansion, $\left(\ln\frac T\Lambda\right)^{-1}$ \cite{hyd3}, which unfortunately is not enough to determine its value at the critical point $T_c$.
%%%%%%%%%%%%%%%%%%%%%%%%%%%%%%%%%%%%%%%%%%%%%%%%%%%%%%%%%%%%%%%%%%%%%%%%%%%%%%%%%%%%%%%%%%%%%%%%%%%%%%%%%%%%%
\begin{figure}[ht]
\begin{center}
\psfrag{zs}{{${\zeta}/{\cS}$}}
\psfrag{tl}{{${T}/{\Lambda}$}}
\includegraphics[width=0.43\textwidth]{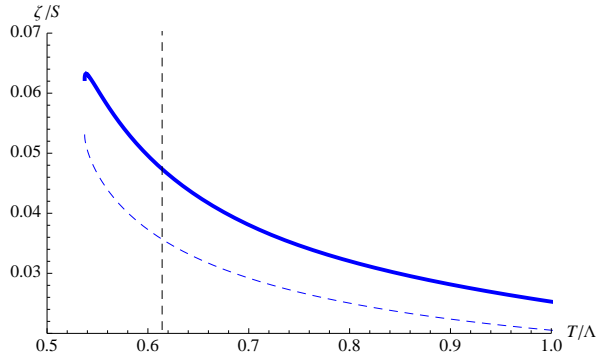}
\end{center}
\caption{(Colour online) The ratio of the bulk viscosity $\zeta$ to the entropy density $\cS$ in cascading gauge theory plasma (solid curve) and the bulk viscosity bound \cite{Buchel:2007mf} (dashed). The dashed vertical line denote the critical temperature $T_c$ of the confinement/deconfinement phase transition.} 
\label{figure1}
\label{fig1}
\end{figure}
%%%%%%%%%%%%%%%%%%%%%%%%%%%%%%%%%%%%%%%%%%%%%%%%%%%%%%%%%%%%%%%%%%%%%%%%%%%%%%%%%%%%%%%%%%%%%%%%%%%%%%%%%%%%%
In turn, we take advantage of the Eling-Oz formula \cite{Eling:2011ms,Buchel:2011yv} to compute the bulk viscosity of the deconfined phase of the cascading gauge theory over all temperature range. The results are presented in Fig.~\ref{figure1}. We find, in particular,
\be
\frac{\zeta}{\cS}\bigg|_{T=T_c}=0.04(8)\,.
\label{bigc}
\ee
Besides, it is worth noticing that the bulk viscosity bound \cite{Buchel:2007mf} is respected all across the phase transition.

We can now address the central question raised in this paper, whether or not cavitation is expected to affect the temperature of the confinement/deconfinement transition in a cascading plasma. Here, the phase $A$ of a fluid is the deconfined phase of the plasma, and $B$ is the confined phase. Since in the planar limit both the transport coefficients and the entropy density are suppressed, we obtain, combining (\ref{big}) and (\ref{bigc}),
\be
\frac{|\delta T_c|}{T_c}\ \lesssim\ \frac{\zeta_A}{\cS_A}=0.04(8)\,.
\ee    
This is an upper bound. Consistency of the hydrodynamics expansion suggests that the left hand side is strictly lower than the right hand side.
   
~

%%%%%%%%%%%%%%%%%%%%%%%%%%%%%%%%%%%%%%%%%%%%%%%%%%%%%%%%%%%%%%%%%%%%%%%%%%%%%%%%%%%%%%%%%%%%%%%%%%%%%%%%%%%%%
\noindent
\textit{Discussion.---}
%%%%%%%%%%%%%%%%%%%%%%%%%%%%%%%%%%%%%%%%%%%%%%%%%%%%%%%%%%%%%%%%%%%%%%%%%%%%%%%%%%%%%%%%%%%%%%%%%%%%%%%%%%%%%
In this Letter we asked to which extent cavitation in confining  gauge theories affects the critical temperature of the confinement/deconfinement transition. We used the specific example of a cascading gauge theory to argue that in the planar limit and at strong coupling the effect is small. It is reasonable to expect that the result is universal as it reflects the fact that large-$N$ phase transitions are typically strong (as opposite to weak) first-order, and that the bulk viscosity at the critical point remains finite.

Some phenomenological models suggest \cite{Karsch:2007jc} that QCD's bulk viscosity might diverge at the critical point of the $T - \mu_B$ phase diagram. Since the QCD critical point \cite{Stephanov:2004wx} separates the line of first-order phase transitions (at large chemical potential) from crossovers (at low chemical potential), both of these effects tend to increase the actual value of $|\delta T_c|/T_c$. This means that hadronization in QCD's expanding quark-gluon fireball, if driven by cavitation, might start earlier than naively expected. Quadratic terms in (\ref{pAB}), though, may become relevant in that case.

~

%%%%%%%%%%%%%%%%%%%%%%%%%%%%%%%%%%%%%%%%%%%%%%%%%%%%%%%%%%%%%%%%%%%%%%%%%%%%%%%%%%%%%%%%%%%%%%%%%%%%%%%%%%%%%
\noindent
\textit{Acknowledgements.---}
%%%%%%%%%%%%%%%%%%%%%%%%%%%%%%%%%%%%%%%%%%%%%%%%%%%%%%%%%%%%%%%%%%%%%%%%%%%%%%%%%%%%%%%%%%%%%%%%%%%%%%%%%%%%%
%
We wish to thank Ofer Aharony, Rob Myers and Aninda Sinha for valuable discussions.
AB is supported by the NSERC through Discovery Grant. AB thanks University of Santiago de Compostela for hospitality while this work was completed. Research at Perimeter Institute is supported through Industry Canada and by the Province of Ontario through the Ministry of Research \& Innovation.
The work of XOC and JDE is supported in part by MICINN and FEDER (grant FPA2011-22594), by Xunta de Galicia (Conseller\'{\i}a de Educaci\'on and grant PGIDIT10PXIB206075PR), and by the Spanish Consolider-Ingenio 2010 Programme CPAN (CSD2007-00042).
XOC thanks the Perimeter Institute for hospitality at the initial stages of this work. He is supported by a Spanish FPU fellowship and thankful to the Front of Galician-speaking Scientists for encouragement.
The Centro de Estudios Cient\'\i ficos (CECs) is funded by the Chilean Government through the Centers of Excellence Base Financing Program of Conicyt.

%%%%%%%%%%%%%%%%%
%%%%%%%%%%%%%%%%%%%%%%%%%%%%%%%%%%%%%%%%%%%%%%%%%%%%%%%%%%%%%
%%%%%%%%%%%%%%%%%
%\bibliographystyle{JHEP}
%\bibliography{library}

\end{document}